\documentstyle[12pt]{article}

\input{tcilatex}

\begin{document}

\title{Spectra of 2D H-donors in magnetic field and their Zeeman splitting.}
\author{Maen Odeh \\
UGRU, UAEU,\\
Al-Ain, Abu Dhabi,\\
UAE.\\
email: maenodeh@emirates.net.ae \and Omar Mustafa \\
Department of Physics, Eastern Mediterranean University\\
G. Magusa, North Cyprus, Mersin 10 - Turkey\\
email:omar.mustafa@emu.edu.tr\\
}
\maketitle

\begin{abstract}
The pseudoperturbative shifted{\small \ - }$l$ expansion technique PSLET
[20-25] is employed to study the spectra of 2D H-donors in magnetic field.
The Zeeman effect is investigated and results are found to be excellent
compared to those in ref.[14].
\end{abstract}

\section{\protect\bigskip Introduction}

\bigskip

The most prominent feature of technology in the last decades was the
realization of low dimensional structures. This achievement urged a huge
body of theoretical and experimental research to study different kinds of
interactions constrained by the new structures [1-14]. Perhaps, one of the
most important and simplest is that of hydrogenic impurity in bulk
semiconductors in a constant magnetic field. In such systems, the energy
scales of both coulombic and magnetic potentials are altered because of the
dielectric function of the medium, $\epsilon \neq 1$, and the effective mass
of the electron $m^{\ast }/m\neq 1$. In semiconductors, typical values of
the effective mass $m^{\ast }$ and dielectric constant $\epsilon $ make the
effective Rydberg $R^{\ast }(=13.6\dfrac{m^{\ast }/m}{\epsilon ^{2}}eV)$
about $10^{3}-10^{4}$ times smaller and the dimensionless magnetic field $%
\gamma =\hbar \omega _{c}/2R^{\ast }$ ( in a fixed magnetic field $B$ )
about $10^{4}-10^{5}$ times larger than those of the free hydrogenic atom
[2,13]. As a result, the situation where $\gamma \sim 1$ ( i.e., the
intermediate magnetic field regime, where neither the Coulombic nore the
confining potentials can be considered small relative to each other ) is of
special physical interest. Therefore, many methods had been employed to
resolve its significance [14].

However, all calculations have shown that the donor states strongly confined
by quantum wells can be correctly described by two dimensional (2D)
hydrongenlike atoms with properly variational parameters [9-12].We can thus
study 2D hydrogenic donors in a magnetic field to understand the behavior of
donors strongly confined by a quantum wells in different magnetic fields.
Although the Hamiltonian models in these settings can be separable, one is
still prevented from obtaining an analytically exact solution. Only for the
two extreme limits of the magnetic field ( $\gamma =0$ and $\gamma
\rightarrow \infty )$ the problem can be treated exactly [1,3]. Recently,
Taut [15,16] has obtained ( conditionally) exact solutions for specific
values of $\gamma $, however no ground-state solutions were obtained.

The adiabatic method [1,3] had been employed in $\gamma \gg 1$ regime.
However, calculations based on this method have usually been restricted to
the ground state and the first few excited states. Other work has been
established using variational method. Makado and McGill [4] have based their
approach on that of Aldich and Greene [2] with different set of basis and
they got quite good results. Martin et al [11] have used a two-point
quasifractional approximation and found excellent interpolation between the
weak and strong magnetic field perturbation expansions. MacDonald and
Ritchie [13] have used a two-point Pade' approximation. The curves they
obtained from different Pade' approximation are very different so that no
regular pattern appears and the results become unreliable [4]. Zhu et al
[14] have used a power series expansion method. Villalbe and Pino [17] have
used a finite-difference scheme with a linear mesh of up to 2000 points and
failed to provide a good estimation of the ground state for the 2D hydrogen
atom. Mustafa [18] and Quigora et al [19] have used the shifted N-expansion
technique and their results were in good agreement with those of Martin et
al [11].

Recently, we have introduced a pseudoperturbative shifted-$l$ ( $l$ is the
angular momentum quantum number ) expansion technique ( PSLET ) to solve Schr%
\={o}dinger equation for states with arbitrary number of nodal zeros. It
simply consists of using $1/\bar{l}$ as a pseudoperturbation parameter,
where $\bar{l}=l-\beta $, and $\beta $ is a suitable shift. The shift $\beta 
$ is vital for it removes the poles that would emerge, at the lowest orbital
states with $l=0$, in our proposed expansion below. Our new analytical,
often semi-analytical, methodical proposal PSLET has been successfully
applied to quasi - relativistic harmonic oscillator [20], spiked harmonic
oscillator [21], anharmonic oscillators [22],  two dimensional hydrogenic
atom in an arbitrary magnetic field [23], two electrons in quantum parabolic
dots [24], truncated coulombic potentials [25], ...etc [26].

Encouraged by its satisfactory performance, we extend, in this paper, PSLET
recipe ( in section 2) and introduce its 2D version to treat the problem of
2D hydrogenic donors in magnetic fields. In section 3, we discuss the
results of PSLET. We conclude our work in section 4.\ \ \ \ \ 

\ \ \ \ \ \ \ \ \ \ \ \ \ \ \ \ \ \ \ \ \ \ \ \ \ \ \ \ \ \ \ \ \ \ \ \ \ \
\ \ \ \ \ \ \ \ \ \ \ \ \ \ \ \ \ \ \ \ \ \ \ \ \ \ \ \ \ \ \ \ \ \ \ \ \ \
\ \ \ \ \ \ \ \ \ \ \ \ \ \ \ \ \ \ \ \ \ \ \ \ \ \ \ \ \ \ \ \ \ \ \ \ \ \
\ \ \ \ \ \ \ \ \ \ \ \ \ \ \ \ \ \ \ \ \ \ \ \ \ \ \ \ \ \ \ \ \ \ \ \ \ \
\ \ \ \ \ \ \ \ \ \ \ \ \ \ \ \ \ \ \ \ \ \ \ \ \ \ \ \ \ \ \ \ \ \ \ \ \ \
\ \ \ \ \ \ \ \ \ \ \ \ \ \ \ \ \ \ \ \ \ \ \ \ \ \ \ \ \ \ \ \ \ \ \ \ \ \
\ \ \ \ \ \ \ \ \ \ \ \ \ \ \ \ \ \ \ \ \ \ \ \ \ \ \ \ \ \ \ \ \ \ \ \ \ \
\ \ \ \ \ \ \ \ \ \ \ \ \ \ \ \ \ \ \ \ \ \ \ \ \ \ \ \ \ \ \ \ \ \ \ \ \ \
\ \ \ \ \ \ \ \ \ \ \ \ \ \ \ \ \ \ \ \ \ \ \ \ \ \ \ \ \ \ \ \ \ \ \ \ \ \
\ \ \ \ \ \ \ \ \ \ \ \ \ \ \ \ \ \ \ \ \ \ \ \ \ \ \ \ \ \ \ \ \ \ \ \ \ \
\ \ \ \ \ \ \ \ \ \ \ \ \ \ \ \ \ \ \ \ \ \ \ \ \ \ \ \ \ \ \ \ \ \ \ \ \ \
\ \ \ \ \ \ \ \ \ \ \ \ \ \ \ \ \ \ \ \ \ \ \ \ \ \ \ \ \ \ \ \ \ \ \ \ \ \
\ \ \ \ \ \ \ \ \ \ \ \ \ \ \ \ \ \ \ \ \ \ \ \ \ \ \ \ \ \ \ \ \ \ \ \ \ \
\ \ \ \ \ \ \ \ \ \ \ \ \ \ \ \ \ \ \ \ \ \ \ \ \ \ \ \ \ \ \ \ \ \ \ \ \ \
\ \ \ \ \ \ \ \ \ \ \ \ \ \ \ \ \ \ \ \ \ \ \ \ \ \ \ \ \ \ \ \ \ \ \ \ \ \
\ \ \ \ \ \ \ \ \ \ \ \ \ \ \ \ \ \ \ \ \ \ \ \ \ \ \ \ \ \ \ \ \ \ \ \ \ \
\ \ \ \ \ \ \ \ \ \ \ \ \ \ \ \ \ \ \ \ \ \ \ \ \ \ \ \ \ \ \ \ \ \ \ \ \ \
\ \ \ \ \ \ \ \ \ \ \ \ \ \ \ \ \ \ \ \ \ \ \ \ \ \ \ \ \ \ \ \ \ \ \ \ \ \
\ \ \ \ \ \ \ \ \ \ \ \ \ \ \ \ \ \ \ \ \ \ \ \ \ \ \ \ \ \ \ \ \ \ \ \ \ \
\ \ \ \ \ \ \ \ \ \ \ \ \ \ \ \ \ \ \ \ \ \ \ \ \ \ \ \ \ \ \ \ \ \ \ \ \ \
\ \ \ \ \ \ \ \ \ \ \ \ \ \ \ \ \ \ \ \ \ \ \ \ \ \ \ \ \ \ \ \ \ \ \ \ \ \
\ \ \ \ \ \ \ \ \ \ \ \ \ \ \ \ \ \ \ \ \ \ \ \ \ \ \ \ \ \ \ \ \ \ \ \ \ \
\ \ \ \ \ \ \ \ \ \ \ \ \ \ \ \ \ \ \ \ \ \ \ \ \ \ \ \ \ \ \ \ \ \ \ \ \ \
\ \ \ \ \ \ \ \ \ \ \ \ \ \ \ \ \ \ \ \ \ \ \ \ \ \ \ \ \ \ \ \ \ \ \ \ \ \
\ \ \ \ \ \ \ \ \ \ \ \ \ \ \ \ \ \ \ \ \ \ \ \ \ \ \ \ \ \ \ \ \ \ \ \ \ \
\ \ \ \ \ \ \ \ \ \ \ \ \ \ \ \ \ \ \ \ \ \ \ \ \ \ \ \ \ \ \ \ \ \ \ \ \ \
\ \ \ \ \ \ \ \ \ \ \ \ \ \ \ \ \ \ \ \ \ \ \ \ \ \ \ \ \ \ \ \ \ \ \ \ \ \
\ \ \ \ \ \ \ \ \ \ \ \ \ \ \ \ \ \ \ \ \ \ \ \ \ \ \ \ \ \ \ \ \ \ \ \ \ \
\ \ \ \ \ \ 

\section{\protect\bigskip \protect\bigskip Hamiltonian model and PSLET recipe%
}

\subsection{\protect\bigskip Hamiltonian model}

\bigskip

The problem of a 2D hydrogenic donor in the presence of a magnetic field
that is perpendicular to the 2D plane, in the effective mass approximation,
can be described by the Hamiltonian [14]

\begin{equation}
\hat{H}=-\dfrac{\hbar ^{2}}{2\mu }\stackrel{\rightarrow }{\nabla _{2}}^{2}+%
\dfrac{e^{2}}{2\mu c^{2}}\stackrel{\rightarrow }{A}.\stackrel{\rightarrow }{A%
}+i\,\dfrac{\hbar e}{\mu c}\stackrel{\rightarrow }{A}.\stackrel{\rightarrow 
}{\nabla _{2}}-\,\dfrac{we^{2}}{\epsilon \rho }
\end{equation}
where $\stackrel{\rightarrow }{A}$ is the vector potential, $\mu $ is the
electron reduced mass, $\epsilon $ is the static dielectric constant, $w$ is
equal to 0 or 1 and $\stackrel{\rightarrow }{\nabla _{2}}$ represents $(%
\hat{\imath}\dfrac{\partial }{\partial x}+\hat{\jmath}\dfrac{\partial }{%
\partial y}).$ If we choose the cylindrical gauge such that 
\begin{equation}
\stackrel{\rightarrow }{A}\,=\dfrac{1}{2}\stackrel{\rightarrow }{B}\times 
\stackrel{\rightarrow }{r}
\end{equation}
and the wavefunction $\Psi (\rho ,\varphi )$ to be in the form 
\begin{equation}
\Psi (\rho ,\varphi )=e^{im\varphi }\dfrac{u(\rho )}{\sqrt{\rho }}
\end{equation}
then the Schr\H{o}dinger like equation that describes the system reads 
\begin{equation}
\left[ -\dfrac{d^{2}}{d\rho ^{2}}+\dfrac{m^{2}-%
{\frac14}%
}{\rho ^{2}}+(\dfrac{\gamma ^{2}}{4}\rho ^{2}-\dfrac{2w}{\rho })\right]
\,u(\rho )=\left[ E(m)-m\gamma \right] \,u(\rho )
\end{equation}
where $m$ is a well-defined magnetic quantum number of the electron in the
cylindrically symmetric magnetic and 2D Coulomb potentials,  $E(m)$ is the
eigenenergy in units of $R^{\ast }=\mu e^{4}/2\hbar ^{2}\epsilon ^{2}$, the
effective Rydberg, $\gamma =\hbar \omega _{c}/2R^{\ast }$, $\omega
_{c}\,(=eB/\mu c)$ is the cyclotron frequency and $\rho $ is in units of the
effective Bohr radius $(a^{\ast }=\epsilon \hbar ^{2}/e^{2}\mu ).$

\subsection{PSLET recipe}

\bigskip

The 2D version of PSLET starts with shifting the magnetic quantum number
through $\widetilde{l}=l-\beta $, where $l=\left| m\right| $, and $\beta $
is a suitable shift introduced technically to avoid the trivial case $l=0,$
and to be determined below. Eq.(4) thus becomes 
\begin{equation}
\left[ -\dfrac{d^{2}}{d\rho ^{2}}+\dfrac{(\widetilde{l}+\beta +1/2)(%
\widetilde{l}+\beta -1/2)}{\rho ^{2}}+\dfrac{\widetilde{l}^{2}}{Q}V(\rho )%
\right] \,u(\rho )={\large \varepsilon }{\Large \,}u(\rho )
\end{equation}
where $\varepsilon =E(m)-m\gamma $, $V(\rho )=\dfrac{1}{4}\gamma ^{2}\rho
^{2}-\dfrac{2w}{\rho }$, and $Q$ is a scaling factor that is set equal to $%
\widetilde{l}^{2}$ at the end of the  calculations .

We employ Taylor's theorem in expanding Eq.(5) about $\rho _{o}$, an
arbitrary point on the $\rho $ axis. It is convenient then to transform the
coordinates in Eq.(5) via $x=\widetilde{l}^{1/2}(\rho -\rho _{o})/\rho _{o}$%
. Expansions about $x=0$ ( i.e., $\rho =\rho _{o}$ ) yield

\bigskip

\begin{equation}
\left[ -\frac{d^{2}}{dx^{2}}+\widetilde{V}(x(\rho ))\right] \,u(x)=\frac{%
\rho _{o}^{2}}{\widetilde{l}}\,{\large \varepsilon \,}u(x),
\end{equation}
where 
\begin{eqnarray}
\widetilde{V}(x(\rho )) &=&\rho _{o}^{2}\widetilde{l}\left[ \frac{1}{\rho
_{o}^{2}}+\frac{V(\rho _{o})}{Q}\right] +\widetilde{l}^{1/2}\left[ -2+\frac{%
V^{^{\prime }}(\rho _{o})\rho _{o}^{3}}{Q}\right] x  \nonumber \\
&+&\left[ 3+\frac{V^{^{\prime \prime }}(\rho _{o})\rho _{o}^{4}}{2Q}\right]
x^{2}+2\beta \sum_{n=1}^{\infty }(-1)^{n}(n+1)x^{n}\widetilde{l}^{-n/2} 
\nonumber \\
&+&\sum_{n=3}^{\infty }\left[ (-1)^{n}(n+1)x^{n}+\left( \frac{d^{n}V(\rho
_{o})}{d\rho _{o}^{n}}\right) \frac{\rho _{o}^{2}(\rho _{o}x)^{n}}{n!Q}%
\right] \widetilde{l}^{-(n-2)/2}  \nonumber \\
&+&(\beta ^{2}-\dfrac{1}{4})\sum_{n=0}^{\infty }(-1)^{n}(n+1)x^{n}\widetilde{%
l}^{-(n+2)/2}+2\beta ,
\end{eqnarray}
and the prime on $V(\rho _{o})$ denotes the derivative with respect to $\rho
_{o}$. It is convenient to expand $\varepsilon $ as

\bigskip

\begin{equation}
{\large \varepsilon }=\sum_{n=-2}^{\infty }E_{n_{\rho },l}^{(n)}\widetilde{l}%
^{-n}..
\end{equation}
\newline
Equation (6) is exactly of the type of Schr\"{o}dinger equation for one -
dimensional anharmonic oscillator. This leads to the identification of the
leading correction in the energy series, Eq.(8), namely\newline
\begin{equation}
E_{n_{\rho },l}^{(-2)}=\frac{1}{\rho _{o}^{2}}+\frac{V(\rho _{o})}{Q}
\end{equation}
\newline
Here $\rho _{o}$ is chosen to minimize $E_{n_{\rho },l}^{(-2)}$, i. e.%
\newline
\begin{equation}
\frac{dE_{n_{\rho },l}^{(-2)}}{d\rho _{o}}=0~~~~and~~~~\frac{d^{2}E_{n_{\rho
},l}^{(-2)}}{d\rho _{o}^{2}}>0,
\end{equation}
\newline
which in turn gives, with $\widetilde{l}=\sqrt{Q}$,\newline
\begin{equation}
l-\beta =\sqrt{\rho _{o}^{3}V^{^{\prime }}(\rho _{o})/2}.
\end{equation}
\newline

Eq.(11) is an explicit equation in $\rho _{o}$. However, closed form
solutions are usually hard to be found, if not impossible. Thus we use
numerical solutions to solve for $\rho _{o}$ ( hence, the technique is often
called semi-analytical ). The shifting parameter $\beta $ is determined by
choosing $\widetilde{l}E_{n_{\rho },l}^{(-1)}=0$. This choice is physically
motivated. It requires not only the agreements between PSLET eigenvalues and
the exact known ones for the harmonic oscillator and Coulomb potentials but
also between the eigenfunctions as well. Hence\newline
\begin{equation}
\beta =-\frac{1}{2}\,(n_{\rho }+\frac{1}{2})\,\Omega ,
\end{equation}
\newline
where\newline
\begin{equation}
\Omega =2\,\sqrt{3+\frac{\rho _{o}V^{^{\prime \prime }}(\rho _{o})}{%
V^{^{\prime }}(\rho _{o})}}.
\end{equation}
Equation (7) thus becomes\newline
\begin{equation}
\widetilde{V}(x(\rho ))=\rho _{o}^{2}\,\widetilde{l}\left[ \frac{1}{\rho
_{o}^{2}}+\frac{V(\rho _{o})}{Q}\right] +\sum_{n=0}^{\infty }v^{(n)}(x)\,%
\widetilde{l}^{-n/2},
\end{equation}
\newline
where\newline
\begin{equation}
v^{(0)}(x)=\frac{1}{4}\,\Omega ^{2}\,x^{2}+2\,\beta ,
\end{equation}
\newline
\begin{equation}
v^{(1)}(x)=-4\beta x-4x^{3}+\frac{\rho _{o}^{5}\,V^{^{\prime \prime \prime
}}(\rho _{o})}{6\,Q}\,x^{3},
\end{equation}
\newline
and for $n\geq 2$\newline
\begin{eqnarray}
v^{(n)}(x) &=&(-1)^{n}\,2\,\beta \,(n+1)\,x^{n}+(-1)^{n}\,(\beta
^{2}-1/4)(n-1)\,x^{(n-2)}  \nonumber \\
&&+B_{n}\,x^{n+2},
\end{eqnarray}

\begin{equation}
B_{n}=(-1)^{n}(n+3)+\frac{\rho _{o}^{(n+4)}}{Q\,(n+2)!}\frac{%
d\,^{n+2}\,V(\rho _{o})}{d\rho _{o}^{n+2}}
\end{equation}
\newline
Equation (6) thus becomes\newline
\begin{eqnarray}
&&\left[ -\frac{d^{2}}{dx^{2}}+\sum_{n=0}^{\infty }v^{(n)}(x)\,\widetilde{l}%
^{-n/2}\right] \,u(x)=\rho _{o}^{2}\left[ \sum_{n=1}^{\infty }E_{n_{\rho
},l}^{(n-1)}\,\widetilde{l}^{-n}\right] \,u(x)  \nonumber \\
&&
\end{eqnarray}
Up to this point, one would conclude that the above procedure is nothing but
an animation of the eminent shifted large-N expansion ( SLNT ). However,
because of the limited capabilities of SLNT in handling large -order
corrections via the standard Rayleigh-Schr\H{o}dinger perturbation theory,
only low-order corrections have been reported, sacrificing in effect its
preciseness. Therefore, one should seek for an alternative and proceed by
setting the wave functions with any number of nodes as

\begin{equation}
u(x)=F_{n_{\rho },l}(x)\exp (U_{n_{\rho },l}(x))
\end{equation}
Eq.(19) readily transforms into the following Riccati type equation 
\begin{equation}
\begin{array}{l}
-\left[ F_{n_{\rho },l}^{^{\prime \prime }}(x)+2F_{n_{\rho },l}^{^{\prime
}}(x)U_{n_{\rho },l}^{^{^{\prime }}}(x)\right] +F_{n_{\rho },l}(x){\Huge \{}-%
{\Huge [}U_{n_{\rho },l}^{^{^{\prime \prime }}}(x)+(U_{n_{\rho
},l}^{^{\prime }}(x))^{2}{\Huge ]} \\ 
\\ 
+2\beta +\dfrac{1}{4}\Omega ^{2}x^{2}+\sum_{n=1}^{\infty }v^{(n)}(x)%
\widetilde{l}^{-n/2}{\Huge \}}=\rho _{o}^{2}\,F_{n_{\rho
},l}(x)\sum_{n=1}^{\infty }E_{n_{\rho },l}^{(n-1)}\widetilde{l}^{-n}
\end{array}
\end{equation}
where primes denotes derivatives with respect to $x$.It is evident that (21)
admits solutions of the form

\begin{equation}
F_{n_{\rho },l}(x)=x^{n_{\rho }}+\sum_{n=0}^{\infty }\sum_{p=0}^{n_{\rho
}-1}a_{p,n_{\rho }}^{(n)}x^{p}\widetilde{l}^{-n/2}
\end{equation}

\begin{equation}
U_{n_{\rho },l}^{^{\prime }}(x)=\sum_{n=0}^{\infty }U_{n_{\rho }}^{(n)}(x)%
\widetilde{l}^{-n/2}+\sum_{n=0}^{\infty }G_{n_{\rho }}^{(n)}(x)\widetilde{l}%
^{-(n+1)/2},
\end{equation}
\newline
where\newline
\begin{equation}
U_{n_{\rho }}^{(n)}(x)=\sum_{m=0}^{n+1}D_{m,n,n_{\rho
}}x^{2m-1}~~~~;~~~D_{0,n,n_{\rho }}=0,
\end{equation}
\newline
\begin{equation}
G_{n_{\rho }}^{(n)}(x)=\sum_{m=0}^{n+1}C_{m,n,n_{\rho }}x^{2m}.
\end{equation}
\bigskip

Substituting equations (22)-(25) into equation (21) implies 
\begin{equation}
\begin{array}{l}
F_{n_{\rho },l}(x){\Huge [}\sum\limits_{n=0}^{\infty }{\Huge (}U_{n_{\rho
}}^{(n)^{^{\prime }}}(x)\widetilde{l}^{-n/2}+G_{n_{\rho }}^{(n)^{^{\prime
}}}(x)\widetilde{l}^{-(n+1)/2}{\Huge )}-\sum\limits_{n=0}^{\infty
}\sum\limits_{m=0}^{n}{\Huge (}U_{n_{\rho }}^{(m)}(x)U_{n_{\rho }}^{(n-m)}(x)%
\widetilde{l}^{-n/2} \\ 
\\ 
+G_{n_{\rho }}^{(m)}(x)G_{n_{\rho }}^{(n-m)}(x)\widetilde{l}%
^{-(n+2)/2}+2U_{n_{\rho }}^{(m)}(x)G_{n_{\rho }}^{(n-m)}(x)\widetilde{l}%
^{-(n+1)/2}{\Huge )+}\sum\limits_{n=0}^{\infty }v^{(n)}\widetilde{l}^{-n/2}
\\ 
\\ 
-\rho _{0}^{2}\sum\limits_{n=1}^{\infty }E_{n_{\rho },l}^{(n-1)}\widetilde{l}%
^{-n}{\Huge ]}-2F_{n_{\rho },l}^{^{\prime }}(x){\Huge [}\sum\limits_{n=0}^{%
\infty }{\Huge (}U_{n_{\rho }}^{(n)}(x)\widetilde{l}^{-n/2} \\ 
\\ 
+G_{n_{\rho }}^{(n)}(x)\widetilde{l}^{-(n+1)/2}{\Huge )]}-F_{n_{\rho
},l}^{^{\prime \prime }}(x)=0
\end{array}
\end{equation}
Obviously, the solution of equation (26) follows from the uniqueness of
power series representation. Therefore, for a given $n_{\rho }$ we equate
the coefficients of the same powers of $\widetilde{l}$ and $n,$
respectively. One can then calculate the energy eigenvalue and
eigenfunctions from the knowledge of $C_{m,n,n_{\rho }},D_{m,n,n_{\rho }}$
and $a_{p,n_{\rho }}^{(n)}$ in a hierarchical manner. Nevertheless, the
procedure just described is suitable for a software package such as MAPLE to
determine the energy eigenvalue and eigenfunction up to any order of the
pseudoperturbation series.

Although the energy series Eq.(8), could appear divergent, or at best,
asymptotic for small $\widetilde{l}$ one can still calculate the eigen
series to a very good accuracy by performing the sophisticated [N,M] Pade'
approximation [25],

\bigskip 
\begin{equation}
P_{N}^{M}(1/\widetilde{l})=\dfrac{\left( P_{0}+P_{1}/\widetilde{l}+\cdot
\cdot \cdot +P_{M}/\widetilde{l}^{N}\right) }{\left( 1+q_{1}/\widetilde{l}%
+\cdot \cdot \cdot +q_{N}/\widetilde{l}^{M}\right) }
\end{equation}
to the energy series, Eq(8). The energy series, Eq(8), is calculated up to $%
E_{n_{\rho },l}^{(8)}/\widetilde{l}^{8}$ by 
\begin{equation}
\varepsilon _{n_{\rho },l}=\widetilde{l}^{2}E_{n_{\rho
},l}^{(-2)}+E_{n_{\rho },l}^{(0)}+\cdot \cdot \cdot +E_{n_{\rho },l}^{(8)}/%
\widetilde{l}^{8}+O(1/\widetilde{l}^{9}),
\end{equation}
and with the $P_{4}^{4}(1/\widetilde{l})$ Pade' approximant it becomes 
\begin{equation}
{\large \varepsilon }_{n_{\rho },l}[4,4]=\widetilde{l}^{2}E_{n_{\rho
},l}^{(-2)}+P_{4}^{4}(1/\widetilde{l}).
\end{equation}
Our method is therefore well prescribed.

\section{Results and Discussion}

\bigskip

For the potential in hand, Eq.(13) reads 
\begin{equation}
\Omega =2\sqrt{2}\sqrt{\dfrac{2\gamma ^{2}\rho _{o}^{3}+2w}{\gamma ^{2}\rho
_{o}^{3}+4w}}
\end{equation}

Eq(11) in turn yields 
\begin{equation}
l+\sqrt{2}(n_{\rho }+\dfrac{1}{2})\sqrt{\dfrac{2\gamma ^{2}\rho _{o}^{3}+2w}{%
\gamma ^{2}\rho _{o}^{3}+4w}}=\sqrt{\dfrac{1}{4}\gamma ^{2}\rho
_{o}^{4}+w\rho _{o}}
\end{equation}

Once Eq.(31) is solved for given values of $\gamma $, $w$, $n_{\rho }$ and $%
l $,\ the coefficients $C_{m,n,n_{\rho }}$, $D_{m,n,n_{\rho }}$ and $%
a_{p,n_{\rho }}^{(n)}$ are obtained in a sequential manner. Then the
eigenvalues and eigenfunctions are calculated in one batch for each value of 
$n_{\rho },$ $l,$ $m$, $\gamma $ and $w$.

It worths mentioning that the method is able to reproduce the exact know
results for both pure Coulombic ( $\gamma =0$ ) and \ pure magnetic ( $w=0$
) interactions. This is evident from Eq.(9). For the pure Coulomb
interaction ( $\gamma =0$ ), Eq.(9) reads 
\begin{equation}
\widetilde{l}^{2}E_{n_{\rho },l}^{(-2)}=-(n_{\rho }+\left| m\right|
+1/2)^{-2},
\end{equation}
while for the pure magnetic interaction ( $w=0$ ), Eq.(9) reads 
\begin{equation}
\widetilde{l}^{2}E_{n_{\rho },l}^{(-2)}=(2n_{\rho }+m+\left| m\right|
+1)\gamma 
\end{equation}
with identically vanishing higher order corrections.

Table 1 shows the results of PSLET for the energy spectra of $1s$, $2p$, $2s$%
, $3d$, $3p$, $3s$ and $4d$ states excluding the Zeeman term $m\gamma $, for
wide range of the dimensionless magnetic field $\gamma .$We have plotted the
quantum levels of 2D donor states without the Zeeman term in Figure 1. It is
evident that as $\gamma $ approaches zero, the quantum levels approach those
of Eq.(32). Our results compare exactly with those obtained by Zhu et al
[14]. It is evidently seen in Figure 1 that the energy of the quantum levels
increases with $\gamma $. Increasing $\gamma $ results in a narrower quantum
well and hence stronger binding energy. It is interesting to note that the
energies of the excited states increases more rapidly with $\gamma $ than
those of the ground states and the energy differences between these states
increase with the applied field.

To study the effect of the Zeeman term, $m\gamma $, we have obtained the
energy spectra for small values of the $\gamma $. These are displayed in
Table 2. Figure 2 shows the results in table 2 including the Zeeman term.
Clearly as the magnetic field is turned on, the states with $m\neq 0$, i.e., 
$2p,3d,3p,4d..$, split because of their negative and positive $m$. As $%
\gamma $ increases continuously, the energy values of positive and negative $%
m$ state close to those of lower and higher $m=0$ states and the levels with
negative $m$ cross those of positive $m$. For example, as $\gamma $ is
turned on the $3d$ states split into $3d_{+}$ and $3d_{-}$ levels. As $%
\gamma $ increases, the $3d_{-}$ state gets closer to the \ $2s$ state and
crosses the $2p_{+}$ state. While the $3d_{+}$ state gets closer to the $3s$
state and crosses the $4d_{-}$ state. Clearly $2p_{-}$ and $3d_{-}$ states
admit minima due to the competition between the Zeeman term and the
parabolic quantum well \ in this weak range of $\gamma .$

Figure 3 shows the effect of the Zeeman term at high magnetic field. Clearly
the energy levels at high magnetic field are accumulated between the
corresponding $m=0$ states and the Landau levels. The order of the levels is 
$1s,2p_{-},3d_{-}$ ... for the first Landau level and $%
2s,2p_{+},3p_{-},4d_{-},..$ for the second Landau level, and so on. As  $%
\gamma $ decreases, the values decreases until the hybrid, mentioned above
occurs.

\section{Concluding remarks}

Using PSLET, we have obtained the energy spectra of the 2D hydrogenic donor
in magnetic field. Calculated results have shown the quadratic effect of the
magnetic field in partially lifting the degeneracy of 2D hydrogenic donor
states. The linear effect of the magnetic field, i.e., the Zeeman term, has
completely lifted the degeneracy causing hybrids between different states.
The order of levels is obtained at high magnetic field. The results obtained
here show exact agreement with those of Zhu et al [12,14]. The
nonperturbative nature of PSLET makes it possible to treat the problem
irrespective of the range of magnetic field. Moreover, we the wavefunctions
obtained by PSLET could be manipulated to be used as trail ones in
variational approaches. The present problem suggests the investigation of
excitons in parabolic quantum dots as they are known to have same behavior
as 2D donors in magnetic field. We believe that all aspects of this problem
could be revealed through PSLET.

\begin{center}
{\huge Figures Captions}
\end{center}

{\Large Figure 1: }The 2D donor energy in effective Rydberg units, excluding
the Zeeman term $m\gamma $, is shown as a function of the normalized
magnetic field $\gamma $ ranging from 0 to 1 ffor $1s,2p,2s,3d,3p,3s$ and $%
4d $ states.

{\Large Figure 2:} The 2D donor energy in effective Rydberg units, including
the Zeeman term $m\gamma $, is shown as a function of the normalized
magnetic field $\gamma $ for $2s,2p,3d,3p$ and $4d$ states.

{\Large Figure 3:} Same as figure 1 but including the Zeeman term $m\gamma .$

\end{document}